\documentclass[fleqn,usenatbib]{mnras}

\usepackage{newtxtext,newtxmath}

\usepackage[T1]{fontenc}

\DeclareRobustCommand{\VAN}[3]{#2}
\let\VANthebibliography\thebibliography
\def\thebibliography{\DeclareRobustCommand{\VAN}[3]{##3}\VANthebibliography}

\usepackage{graphicx}	
\usepackage{amsmath}	

\usepackage{amssymb}	

\usepackage{xcolor}

\title[Non-linear waves due to merging flux ropes]{Oscillatory reconnection and waves driven by merging magnetic flux ropes in solar flares}

\author[J. Stewart et al.]{
J. Stewart,$^{1}$\thanks{E-mail: james.stewart@manchester.ac.uk}
P. K. Browning,$^{1}$
M. Gordovskyy$^{1}$
\\
$^{1}$Jodrell Bank Centre for Astrophysics, University of Manchester, Manchester M13 9PL, United Kingdom\\
}

\date{Accepted XXX. Received YYY; in original form ZZZ}

\pubyear{2022}

\begin{document}
\label{firstpage}
\pagerange{\pageref{firstpage}--\pageref{lastpage}}
\maketitle

\begin{abstract}

Oscillatory reconnection is a process that has been suggested to underlie several solar and stellar  phenomena, and is likely to play an important role in  transient events such as flares. Quasi-periodic pulsations (QPPs) in flare emissions may be a manifestation of oscillatory reconnection, but the underlying mechanisms remain uncertain.  In this paper, we present 2D magnetohydrodynamic (MHD) simulations of two current-carrying magnetic flux ropes with an out-of-plane magnetic field undergoing oscillatory reconnection in which the two flux ropes merge into a single flux rope.  We find that oscillatory reconnection can occur intrinsically without an external oscillatory driver during flux rope coalescence, which may occur both during large-scale  coronal loop interactions and the merging of plasmoids in fragmented current sheets. Furthermore, we demonstrate that radially propagating non-linear waves are produced in the aftermath of flux rope coalescence, due to the post-reconnection oscillations of the merged flux rope.  The behaviour of these waves is found to be almost independent of the initial out-of-plane magnetic field. It is estimated that the  waves emitted through merging coronal loops and merging plasmoids in loop-top current sheets would have a typical  phase speed of 90 km/s and 900 km/s respectively. It is possible that the properties of the waves emitted during flux rope coalescence could be used as a diagnostic tool to determine physical parameters within a coalescing region.

\end{abstract}




\section{Introduction}

Solar flares are characterised by the short-term brightening of emissions from the solar atmosphere across the electromagnetic spectrum \citep{Fletcher2011,Benz2017}. The duration of solar flares can range from several minutes to several hours and are associated with brightened emissions in the radio, UV/EUV, X-ray and $\gamma$-ray bands. Stellar flares have been observed in many types of star, including powerful variants referred to as ``superflares", such as those reported in \citet{Tu2020}. In comparison to solar flares, stellar flares have been observed to last up to several days, however their emissions have been observed to occupy similar frequency bands to those emitted by solar flares \citep{Benz2017}.\\

Theoretical models of solar flares must be able to account for this wide range of enhanced emissions. It is widely accepted that magnetic reconnection is the energy release mechanism that leads to these brightened emissions \citep{Shibata2011}. However, there are several scenarios in which magnetic reconnection can occur in the solar atmosphere. These scenarios typically involve either the interaction of two coronal loops (two-loop models) or the interaction of a coronal loop with the solar magnetic field (one-loop models). Examples of one-loop models include the CSHKP model of a flaring filament, developed by the combined efforts of \citet{Carmichael,Sturrock,Hirayama1974,Kopp} and the interaction of an emerging coronal loop with its surrounding magnetic field, developed by \citet{Antiochos98}. Examples of two-loop models include two non-parallel loops interacting with each other, as described by \citet{SweetA,Kumar2010, Li2021}, and an emerging coronal loop interacting with a preexisting loop in the solar atmosphere, as described by \citet{Heyvaerts, Zheng_2018, Zhong_2019}. Two-loop models are one focus of this paper, as are fragmentary twisted structures within single-loop configurations. \\

The aforementioned models predict well the general behaviour of emission brightening in solar flares. However, they do not account for temporary quasi-periodic oscillations, which are frequently observed in solar flare emissions. Observations of these oscillations, referred to as quasi-periodic pulsations (QPPs), can be traced back to a review of solar continuum radio bursts by \citet{Thompson1962}. QPPs however were not the focus of this review. Attention was first drawn to QPPs in a paper by \citet{Parks1969-wy}; in which the authors discuss a sixteen-second periodic modulation in the X-ray intensity-time profile of a 1968 solar flare. The current understanding of QPPs in both solar and stellar flares is summarised in recent reviews by  \citet{McLaughlin2018, Nakariakov2019, Doorsselaere2020, QPPReview2021}.  QPPs are manifest in flare light curves across the electromagnetic spectrum from radio to gamma-rays, and are observed to have a wide variety of periods, which may be due to different propagation or driving mechanisms (\citet{QPPReview2021}). They are also observed to have a wide variety of temporal behaviours, including:  QPPs with aperiodic trends, anharmonic shape, modulated period and amplitude and those superimposed with background noise. Recent advances in techniques for robust detection of solar and stellar QPPs are outlined by \citet{Broomhall2019}.\\

Observations of QPPs are not rare, as revealed by several statistical studies on the prevalence of QPPs within solar flares. \citet{Inglis2016-ko} studied all GOES M and X-type flares between 2011 and 2016 and found signatures of QPPs in the SXR band in 30\% of flares out of a total of 675. \citet{Dominique2018} looked at all solar flares within Solar Cycle 24 with a GOES magnitude of M5 or higher, using wavelet analysis and a set of detection criteria, developed to prevent false observations due to detrending and windowing. They reported that 90\% of the 90 considered flares contained QPPs within the EUV or SXR bands. Further statistical studies indicate that QPPs typically last between a few seconds to a few minutes \citep{QPPReview2021},  though shorter \citep{Takakura1983-jf} and longer events have been observed, with one detection by \citet{Zaqarashvili2013-qv} lasting over 30 minutes. QPPs have also been detected within stellar flares \citep{Mitra2005,Mathioudakis2003,Mathioudakis2006} and in pre-main sequence star flares (for example, two 3-hour period oscillations were observed in the Orion star-forming region by the \textit{Chandra} Orion Ultradeep Project and were successfully modelled by \citet{Reale2018}).\\

The presence of QPPs within flare emissions is important, as it implies the existence of additional physical mechanisms within a solar flare. Furthermore, understanding the mechanisms driving the generation of QPPs could lead to the development of diagnostic tools that could be used to determine the physical parameters within a flaring region. While quantitative models of solar flares are beginning to emerge \citep{Ruan2020}, they are mainly concerned with forward-modelling a plasma with known initial conditions and not the inverse. Studying QPPs would provide a great deal of insight into how the physical conditions inside a solar flare can be determined from observation data. QPPs are a signature of the intrinsically transient energy release in solar flares and thus gaining a better understanding of time-dependent reconnection would allow for better modelling of QPPs and conversely, QPPs may shed light on the time-dependent reconnection process.\\

Despite the wealth of observations, the mechanisms underlying QPPs remains uncertain, with multiple models proposed. \citet{McLaughlin2018} discuss a total of 11 debated QPP mechanisms. In recent years the number of proposed mechanisms has increased, with 13 being reported by \citet{Kupriyanova2020-lp}, and 15 by \citet{QPPReview2021}. Broadly, mechanisms can be categorised as "MHD oscillations" (associated with some MHD waves from an external source) and "self-oscillations" \citep{Nakariakov2019}, but this  has also been a matter of debate. For example, as pointed out by \citet{QPPReview2021}, some mechanisms could fit into more than  one proposed categories. On the other hand, some proposed mechanisms may not be distinct  from each other. One of the difficulties is that many  theoretical QPP models are qualitative and do not account for features such as particle acceleration mechanisms that are important in flare emissions \citep{QPPReview2021}. Furthermore, observations of QPPs do not yet provide a level of detail on the physical properties within flare regions that would allow these questions to be discussed further.\\

There are however, promising lines of research that could work to close this gap between observations and simulations; which  include the study of oscillatory reconnection and of magnetohydrodynamic (MHD) waves generated by magnetic reconnection. QPPs have a very broad range of periods, but for many observations, these are compatible with the expected periods of MHD waves  - although in other cases, there is a discrepancy \citep{Nakariakov2019}.  Furthermore, observations  \citep[e.g.]{Nishizuka2008-wh, He_2009} have discovered waves generated from magnetic reconnection within the solar atmosphere. Further understanding of reconnection-driven MHD waves and pulsations would also allow for the development of diagnostic tools that could be used to determine the physical parameters within a flaring region. For example,  2D MHD simulations by \citet{McLaughlin_2012} demonstrated that oscillatory magnetic reconnection during the emergence of a flux rope into the solar corona can produce MHD waves, with periodic production of vertical outflows with a speed of 20-60 km/s and a periodicity of 1.75-3.5 minutes. \citet{McLaughlin_2012} compared these values with observations of vertical outflows within the quiet and active Sun with a line-of-site speed of 50-150 km/s \citep{DePontieu2009, McIntosh2009} and observations by \citet{McIntosh2010} who measured a high-speed vertical outflow of mean speed 135 km/s with a periodicity of 300-1500 seconds.\\

Further studies continued to develop the connection between oscillatory reconnection and the production of MHD waves. \citet{Thurgood2017} performed 3-dimensional simulations of a collapsing null point due to interactions from an incoming external MHD wave and discovered that the null point entered a state of oscillatory reconnection which produced MHD waves that propagated away from the reconnection site. \citet{karampelas2021oscillatory} extended this model to a hot plasma to better suit the conditions within the solar atmosphere. In doing so, they also provided evidence that oscillatory reconnection could be used as a seismological tool to calculate the physical parameters within a flaring region by finding a relationship between the equilibrium magnetic field strength, and the period and decay rate of oscillatory reconnection oscillations within this system. This result suggests that with further research, that it may be possible to calculate the equilibrium magnetic field strength of an collapsing X-point from the periodicity of observed QPPs. \\

Reconnection in 2D requires the presence of magnetic X-points, and the interaction between waves and X-points - as well as 3D null points -  has been extensively studied (e.g. \citet{McLaughlin2011}), showing that wave energy is preferentially dissipated near the null points. The aforementioned research however, is limited to studying oscillatory reconnection initiated by an incoming MHD wave of  unspecified origin. It is important to determine whether oscillations can arise intrinsically in a reconnecting field. To this end, \citet{Smith2022}, using a realistic model configuration developed by \citet{Gordovskyy2014}, demonstrated that quasi-periodic oscillations could be produced without an external source. However, many open questions remain, including whether oscillatory reconnection at a single site can develop intrinsically, and whether the properties of such a system are consistent with QPP observations.\\

\citet{Stanier2013} present a model of the start-up mechanism in the Mega-Ampere Spherical Tokamak (MAST), providing  a promising  scenario in which  oscillatory reconnection can occur without an exterior driver. They performed 2D MHD and Hall-MHD simulations of magnetic reconnection originating from two current-currying magnetic flux ropes. They found that the flux ropes attracted to each other via the Lorentz force and merged into a single flux rope. Though these simulations were performed in the context of MAST, two factors indicate that this research is applicable to coronal physics. The first is that the merging-compression start-up mechanism of MAST is performed with a strong magnetic field, low-plasma beta and high Lundquist number,  similar to the  solar corona. The second is the use of magnetic flux ropes, which are common structures in the solar corona, often associated with solar flares (e.g. \citet{vemareddy2022eruption, Liu2020, agapitov2021flux}). In this paper, we develop this scenario to investigate  merging flux ropes in the corona, and consider the  implications for flares and QPPs.\\

The interaction and merger of magnetic flux ropes has long been considered to play a role in solar flares \citep{Tajima1987,Sakai1997}, and in QPP production. For example, \citet{Tajima1987} investigate  the nonlinear coalescence  instability between two flux ropes with parallel currents, and find the simulation results  replicate  observed QPPs in the HXR, $\gamma$-ray, and microwave band for the 1980 ``seven-sisters" solar flare.\\

On the smaller scale, large-scale  current sheets are prone to fragmentation into islands or plasmoids if the Lundquist number is sufficiently large \citep{Loureiro2007, Shibata2016}  - a criterion easily met in the solar corona. In the presence of a guide field (out-of-plane field component) these plasmoids are twisted flux ropes. Current sheet fragmentation and subsequent merger of plasmoids has been widely studied with application to both laboratory \citep{Daughton2011} and solar \citep{Barta2011, Karlicky2012,Potter2019} plasmas, and may enable  fast magnetic reconnection. Several observational studies confirm the presence of  plasmoids in reconnecting current sheets in flares \citep{Takasao_2011,Lu2022}. The  model we present here can be scaled to match either the merging of large scale coronal loops or smaller structures within a current sheet. It should be noted that flux ropes mutually attract only if they carry  non-zero net current. This remains a matter of longstanding debate \citep{Melrose1991}, but there is a considerable body of evidence for some coronal flux ropes with non-neutralised current  \citep{Georgoulis2012,Torok2014, Dalmasse2015,Vemareddy2019,He2020}. \\

%
Previous studies such as those performed by \citet{Tam2015} and \citet{Threlfall2018} demonstrate that merging magnetic flux ropes can lead to dynamic processes. However, these studies were concerned with merging flux ropes, with zero-current, triggered by the kink instability of one flux rope, while we consider stable flux ropes which merge due to the mutual attraction of parallel currents. Furthermore, \citet{Tam2015} and \citet{Threlfall2018} consider MHD evolution and particle acceleration respectively, while neither consider the nature of oscillations within the reconnection environment, the possible generation of MHD waves through oscillatory reconnection or the relation of these waves to QPPs. Our goal here is to pursue this, by adapting the flux rope coalescence simulation of \cite{Stanier2013} (which is known to produce oscillatory reconnection) to the solar context.  Whereas \citet{Stanier2013} focus only on the dynamics  of reconnection  during the merging-compression process, we consider also the  surrounding environment where emitted MHD waves could propagate. It is not clear whether the oscillatory reconnection is influenced by the boundary conditions, and we therefore use open boundary conditions rather than rigid conducting walls as used in \citet{Stanier2013}. Furthermore, \citet{Stanier2013} use a uniform resistivity corresponding to the conditions in MAST, but in the solar corona where the Lundquist number is much higher, it is expected that reconnection involves a current-driven anomalous resistivity associated with kinetic instabilities \citep{Singh2007} which we include in our simulations; this may significantly affect the reconnection rate and dynamics \citep{Biskamp1980,Yokoyama1994,Nakariakov1999}.\\

We thus aim to investigate the onset of oscillatory reconnection in merging magnetic flux ropes in flares, and whether this results in the emission of MHD waves without excitation from an exterior source.  To this end, we perform 2D MHD simulations of two merging flux ropes using open boundary conditions and anomalous resistivity, with a focus on identifying the oscillations and waves which arise and exploring how they are related to the reconnection process. 
 Section 2 describes  the flux rope model and its implementation within 2D resistive MHD simulations. Results are presented in Section 3 and discussed in Section 4, with a focus on the physical properties of the emitted waves and their implications for flares and QPPs.

\section{Model and Simulation Description}
\subsection{The Resistive MHD Equations}

We simulate two merging magnetic flux ropes in conditions that represent the coronal environment. To model the flux ropes, we use a modified version of the Lagrangian form of the resistive MHD equations, including an additional viscous force term to account for weak shocks $\bf{f}_{visc}$. This artificial viscosity term was initially developed by \citet{Caramana1998} and adapted for use in MHD by \citet{LareManual}. It contributes a value roughly on the order of magnitude of the difference in the kinetic energy density of a plasma element at a grid-point and its nearest-neighbours. The term is calculated by approximating the fluid as a series of finite volume masses distributed across a staggered grid and calculating the non-linear energy exchange due to inelastic collisions between said particles; a method originating from \citet{VonNeumann1950}. A second linear contribution, developed by \citet{Landshoff1955} is then added to remove nonphysical oscillations that can arise behind a shock front. \citet{Caramana1998} discussed that treating an arbitrarily divided continuous fluid as finite volume masses can lead to errors, such as incorrect viscous dissipation calculations due to self-similar isentropic compression. To account for this, \citet{Caramana1998} introduced a limiter function, adapted from the work of \citep{Christensen1990, Benson1993}, that switches off the artificial viscosity term when the velocity field is a linear function of spatial coordinates, i.e. in smooth regions of flow. Further details of how $\bf{f}_{visc}$ is calculated can be found in \citet{LareManual}.\\

The adapated Lagrangian form of the resistive MHD equations are presented below. All variables in the below equations, and the results presented in this paper are normalised. The normalisation constants used in this paper are defined in Table \ref{Tab:normalisation_constants}. The normalisation constants: $L_0$, $B_0$ and $\rho_0$ are user-defined. The value of these constants for two examples relevant to this paper have been provided - the coronal conditions and length scale of a plasmoid in a fragmented current sheet and that of coronal loop.

\begin{gather}
    \frac{D\rho}{Dt} + \rho\bf{\nabla}\cdot\bf{v} = 0\\
    \rho\frac{D\bf{v}}{Dt} = (\bf{\nabla}\times\bf{B})\times\bf{B} - \bf{\nabla}P + \bf{f}_{visc}\\
    \frac{D\bf{B}}{Dt} = (\bf{B}\cdot\bf{\nabla})\bf{v} - \bf{B}(\bf{\nabla}\cdot\bf{v}) - \bf{\nabla}\times(\eta\bf{\nabla}\times\bf{B})\\
    \frac{D\epsilon}{Dt}  = -\frac{P}{\rho}(\bf{\nabla\cdot\bf{v}}) + \frac{\eta}{\rho}j^2\\
    P = \rho\epsilon(\gamma-1)
\end{gather}

\begin{table*}
\makebox[\textwidth][c]
\centering
\caption{Definition of the normalisation constants used within our simulations. Example values for the normalisation constants that apply to plasmoids in a fragmented sheet and to coronal loops are provided.}
\label{Tab:normalisation_constants}
\begin{tabular}{|c|c|c|c|}
\hline
Normalisation Constant & Definition & Value (Plasmoid) & Value (Coronal Loop) \\ \hline
 $L_0$ & User-Defined & $10^4$ m& $10^6$ m \\ [1.5ex]
  $B_0$& User-Defined & $5*10^{-3}$ T & $5*10^{-3}$ T \\ [1.5ex]
 $\rho_0$ & User-Defined & $10^9$ cm\textsuperscript{-3} &  $10^9$ cm\textsuperscript{-3} \\ [1.5ex]
 $v_0$& $\frac{B_0}{\sqrt{\mu_0\rho_0}}$ &  $3.15*10^6$ ms\textsuperscript{-1}&  $3.15*10^6$ ms\textsuperscript{-1} \\ [1.5ex] 
 $P_0$ & $\frac{B^2_0}{\mu_0}$ &  $1.25*10^7$ Pa & $1.25*10^7$ Pa \\ [1.5ex]
 $t_0$ & $\frac{L_0}{v_0}$  & $3.17*10^{-3}$ s & $0.317$ s \\ [1.5ex]
 $j_0$ & $\frac{B_0}{\mu_0L_0}$ & $0.397$ Am\textsuperscript{-2} & $3.97*10^{-3}$ Am\textsuperscript{-2} \\ [1.5ex]
 $\epsilon_0$ & $v^2_0$ & $9.92*10^{12}$ Jkg\textsuperscript{-1} & $9.92*10^{12}$ Jkg\textsuperscript{-1}  \\ [1.5ex]
 $T_0$ & $\frac{\epsilon_0\Bar{m}}{k_B}$ & $1.44*10^{9}$ K &  $1.44*10^{9}$ K \\ [1.5ex]
 $\eta_0$ & $\mu_0L_0v_0$ & $3.97*10^{4}$ $\Omega$m & $3.97*10^{4}$ $\Omega$m \\ [1.5ex] \hline
\end{tabular}
\end{table*}

The variables in the equations above are defined as follows: $\rho$ is the mass density, $\bf{v}$ is the plasma velocity, $\bf{B}$ is the background magnetic field, $P$ is pressure, $\eta$ is magnetic resistivity, $\epsilon$ is the specific internal energy density, $j$ is the current density and $\gamma$ is the heat capacity ratio. The effects of thermal conduction and radiation have not been included in these simulations. Though energy losses and energy transport have been shown to dampen propagating MHD waves in linear theory \citep{Wang2021}, our focus is on the generation of oscillations and how they relate to the reconnection process. This is not expected to change with the inclusion of heating and energy losses. Nevertheless it should be noted that this limits the accuracy of the observational predictions made by these simulations.

\subsection{Initial and Boundary Conditions}

We now consider the initial conditions. Flux ropes, for small enough length scales, can be approximated as infinitely-long cylinders. We adopt the initial configuration used by \citet{Stanier2013} to calculate the initial magnetic field due to the presence of two twisted magnetic flux ropes and adapt it for use in the solar corona.\\

Consider a single cylindrical flux rope in cylindrical polar coordinates, extending along the z-axis, with a radius $r = w$. The azimuthal angle in this coordinate system is represented by $\phi$. Following \citet{Stanier2013}, the current density along the z-axis for a flux rope in this configuration is taken to be: 

\begin{equation}
    j_z = \left\{
        \begin{array}{ll}
            j_m\left( 1 - \frac{r^2}{w^2}\right )^2, & r \leq w;  \\
             0, &  r > w,
        \end{array}
    \right.
\end{equation}

where at the origin, the current density peaks with a value $j_m$ and then approaches zero as $r \rightarrow w$.\\

The azimuthal magnetic field of the flux rope can then be calculated using Amp\`ere's Law, and is given by

\begin{equation}
    B_\phi = \left\{
        \begin{array}{ll}
            B_p \left ( \frac{3r}{w} - \frac{3r^3}{w^3} + \frac{r^5}{w^5}  \right  ), &  r \leq w ;\\ \\
      
      B_p\frac{w}{r}, &  r > w,
        \end{array}
    \right.
\end{equation}

where $B_p = \frac{wj_m\mu_0}{6}$ is a constant that characterises the magnitude of the poloidal magnetic field. The peak poloidal field is $B_{\phi} = 1.24B_p$.\\

Each flux rope is initially considered to be force-free before being placed in the same environment as the other flux rope. Using this condition, the magnetic field along the z-axis for a magnetic flux rope can be determined by solving $\bf{j}\times\bf{B} = 0$. For $r\leq w$, this gives:

\begin{equation}
    B_z = 
            B_0 \left ( 1 + \frac{B^2_p}{B^2_0}\left ( \frac{47}{10} - \frac{18r^2}{w^2} + \frac{27r^4}{w^4} - \frac{20r^6}{w^6} + \frac{15r^8}{2w^8} - \frac{6r^{10}}{5w^{10}} \right ) \right )^{\frac{1}{2}}, 
\end{equation}

and for $r>w$, $B_z = B_0$, where $B_0$ is the constant value of $B_z$ outside of the flux rope. Due to the symmetry of the system, $B_r = 0$.\\

We consider an initial state of two magnetic flux ropes with width $w = 1.0$ and poloidal field constant $B_p =  1.0$, in a plasma with initial velocity $\bf{v = 0}$ and plasma density $\rho = 1.0$ everywhere. By superposing two flux ropes, we obtain a  magnetic field which is  not force-free; the flux ropes maintain their identities as the ``internal" Lorentz force vanishes but the unbalanced overall Lorentz force (associated with the non-zero currents) causes the ropes to move towards each other \citep{Stanier2013}.  The initial separation  distance along the y-axis  of the flux rope centres is set as   $h = 3.0$ providing a suitable distance in which the two flux ropes do not initially overlap, but close enough that the attractive force between them will quickly pull them both together. The initial temperature is set as $T = 6.7*10^{-3}$, giving a low plasma-beta. The initial  background axial magnetic field is set as  $B_0 = 2.0$, giving an Alfv\'en time of $\tau_{A} = 0.5$, for our main set of simulations, but we also investigate the effects of varying this quantity, where the Alfv\'en time is 1 and $\frac{1}{3}$ for $B_0  = 1$ and $3$ respectively. The ratio of the peak azimuthal field to axial field is thus 0.62, which is appropriate for a strongly-twisted coronal flux rope.\\

Zero-gradient boundary conditions are used for all variables at the edges of the 2D grid, in order to model an open system as close as possible. It should be noted these conditions do allow flows of energy in or out of the system, so total energy is not conserved.

\subsection{Evolution of the System using LARE2D}

We use LARE2D code developed by \citet{Arber2001}, which is a series of Lagrangian remap codes to solve the resistive MHD equations in two-dimensions. Studies such as those done by \citet{McLaughlin_2012} and \citet{Thurgood2017} have shown that LARE2D can be used to successfully simulate oscillatory reconnection.\\

The code defines variables on a staggered Cartesian grid, with a user-defined resistivity and viscosity. These variables are then evolved over time using finite-difference methods to solve the resistive MHD equations. Further description of the code can be found in the LARE manual \citep{LareManual}.\\

The simulation domain used is a 8x8 length box with 1024 x 1024 grid points. The large domain size was chosen to reduce boundary effects as much as possible. Simulations were run until $t = 100$, by which time the system was close to an equilibrium state consisting of a single flux rope. The linear viscosity of the fluid was set to $\nu = 10^{-4}$, removing any significant energy loss due to viscous forces, thus better simulating the coronal environment.\\

As discussed above, a current-driven anomalous resistivity model is commonly used for coronal reconnection simulations. Here, we set the resistivity  to $\eta = 10^{-5}$,  increasing to $\eta = 10^{-3}$ in regions where the current density exceeds a critical value $j_{crit} = 1.2$. This gives a characteristic Lundquist number value of order $S = 10^{5}$ for the background plasma, decreasing to $S = 10^{3}$ in a reconnecting region. The value of the critical current is selected to be larger than the peak of the initial currents, so that dissipation is mainly associated with ``current sheets" rather than background current.

\begin{figure*}
\makebox[\textwidth][c]{\includegraphics[width=1.2\textwidth]{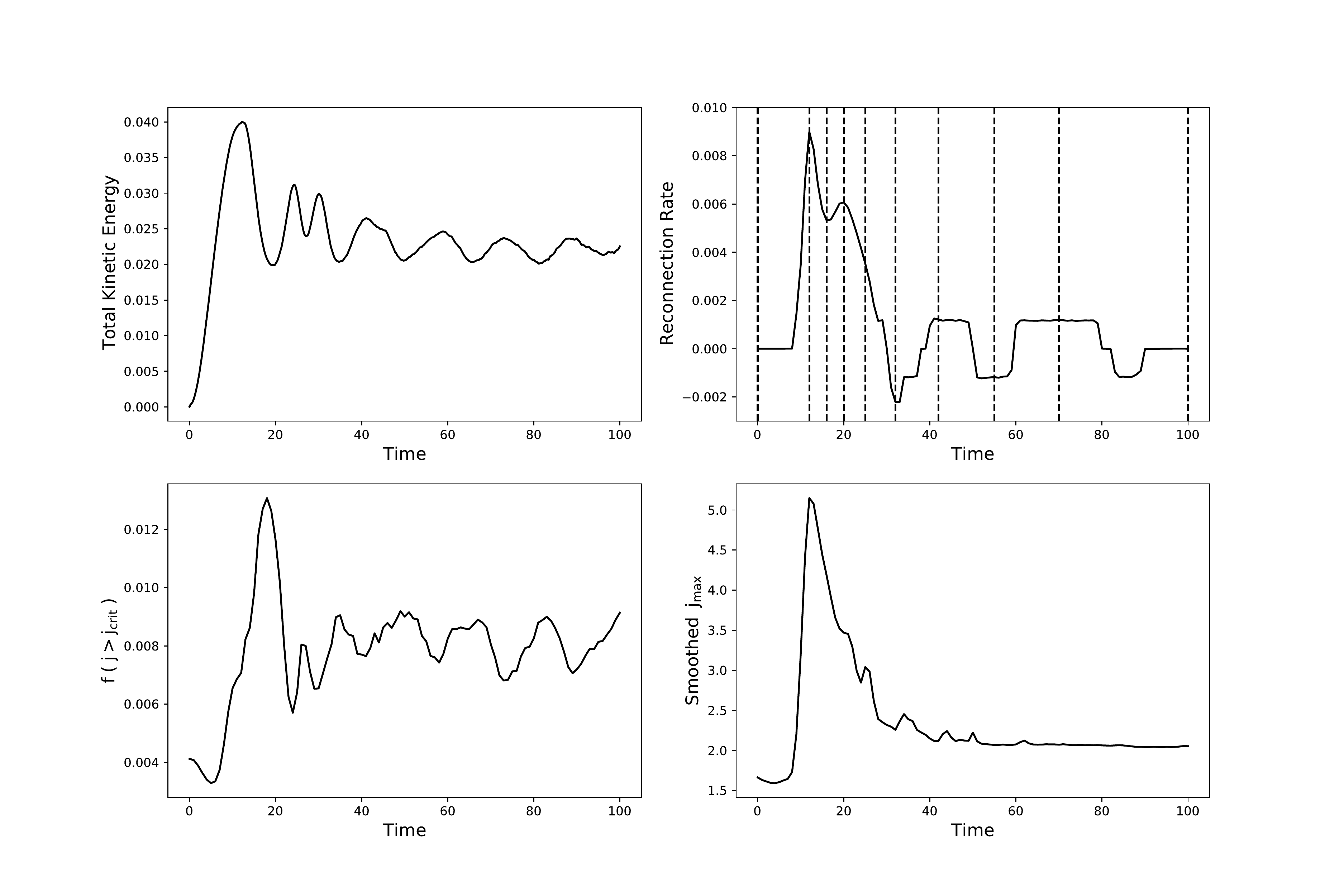}}%
    \caption{Plots of the evolution of the total kinetic energy of the system (top-left), the reconnection rate at the origin (top-right), the fraction of the domain with total current above the critical current (bottom-left) and the smoothed maximum current (bottom-right) are plotted over time. The dashed vertical lines in the reconnection rate plot mark the time in which the snapshots presented in Figure \ref{fig:reconnection_snapshots} and Figure \ref{fig:oscillation_snapshots} are taken.}
    \label{fig:energy_evolution}
\end{figure*}

\begin{figure*}
\makebox[\textwidth][c]{\includegraphics[width=1.2\textwidth]{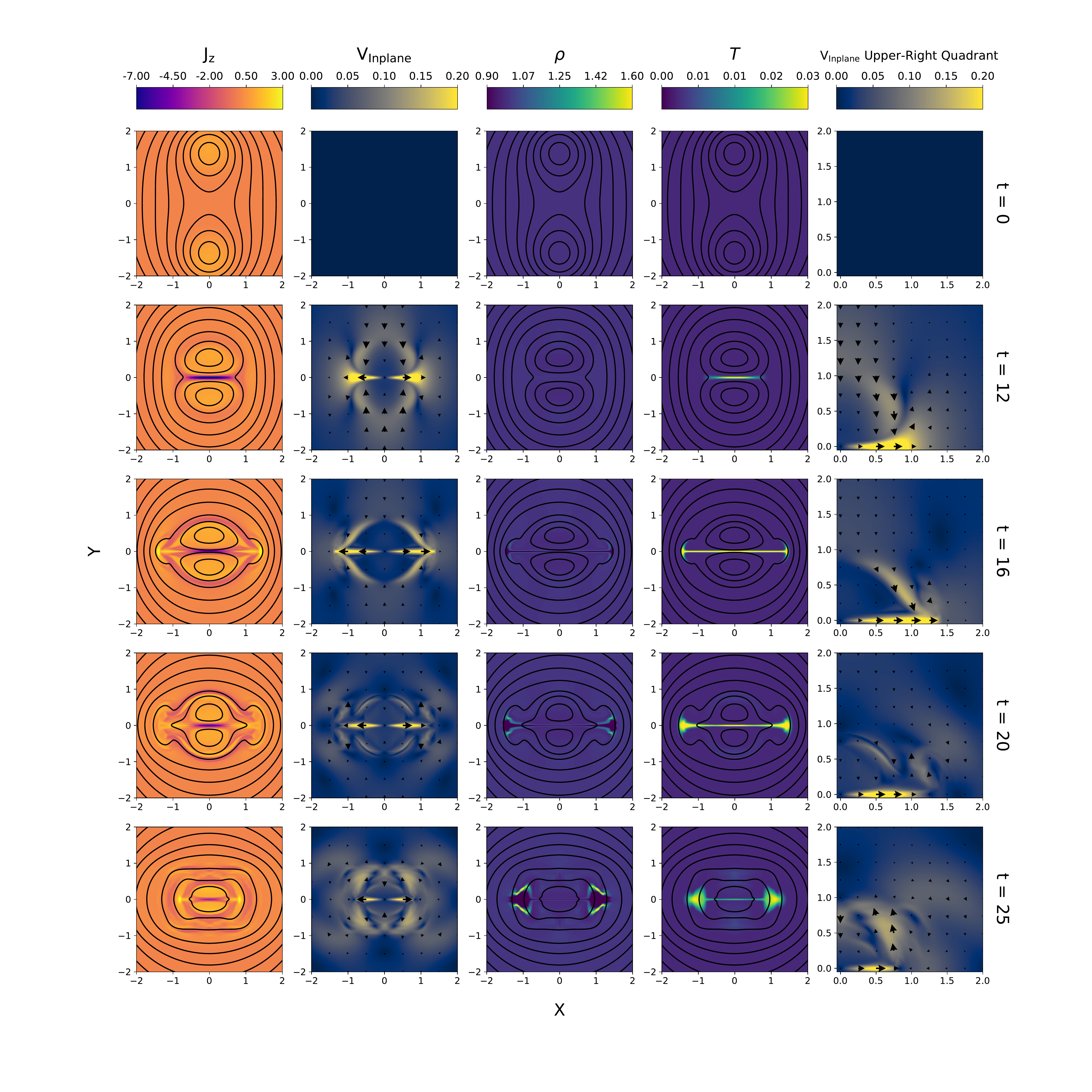}}%
\caption{Colour maps of $J_z$, $V_{in-plane}$, $\rho$ and $T$ at $t = 0, 12, 16, 20, 25$. Each column in the figure corresponds to a different variable, except for column 5, where the plots are zoomed in colour maps of the upper-right quadrant of the respective plots in column 2. Each row corresponds to a different snapshot in time. The time of the snapshots presented in this figure coincide with dashed lines found in the reconnection rate plot in Figure \ref{fig:energy_evolution}. Contour lines of $A_z$ are plotted over $J_z$, $\rho$ and $T$, illustrating the in-plane magnetic field. Scaled arrows have been applied on top of the in-plane velocity colour maps, illustrating the direction and magnitude of the velocity flow. A video of the evolution of these variables between $t = 0 - 100$ can be as supplementary material.}
    \label{fig:reconnection_snapshots}
\end{figure*}

\section{Results}

We simulate two current-carrying magnetic flux ropes undergoing magnetic reconnection in a coronal plasma using LARE2D. The results of these simulations are presented below.\\

The overall time evolution of the system is summarised in Figure \ref{fig:energy_evolution}, in which the evolution of the total kinetic energy, the reconnection rate at the origin, the fraction of the domain with current above the critical current and the smoothed maximum current is shown, providing context for the understanding of the evolution of the magnetic field and other plasma parameters shown in Figures \ref{fig:reconnection_snapshots} and \ref{fig:oscillation_snapshots}. The fraction of the domain with current above the critical current is illustrative of how much of the domain is undergoing reconnection, while the smoothed maximum current illustrates how strong the current sheets in the reconnection process are. The smoothed maximum current is calculated by averaging the top $0.1\%$ of the values of $|\bf{j}|$ within the domain to avoid spurious fluctuations to selecting the highest value on a finite grid.\\

The reconnection rate is calculated as the out-of-plane electric field at the X-point, which defines the rate of flux change in a 2D configuration, invariant in the out-of-plane direction \citep{Stanier2013,Comisso_2016}. In our simulations, the initial X-point is located at the origin and remains there as reconnection commences due to the symmetry of the system. Therefore we calculate the reconnection rate as $-E_z$ at the origin. We choose the minus-sign convention so that the reconnection rate is positive when reconnection begins.\\

In Figures \ref{fig:reconnection_snapshots} and \ref{fig:oscillation_snapshots}, we use contours of the out-of-plane vector potential $A_z$ to illustrate the in-plane magnetic field lines. For a 2D system invariant in z, the magnetic field in the x-y plane can calculated using the vector potential: $\mathbf{A(x,y)} = A_z(x,y)\mathbf{\hat{z}}$. It can be shown that as the tangent vectors of the equipotential contours of $A_z$ are equivalent to the equations of the in-plane magnetic field lines that the in-plane magnetic fields lines can be illustrated by plotting contours of $A_z$.\\

It is convenient to consider the evolution of the system in two phases, discussed further below: an initial merger phase, in which  the flux ropes approach and merge into a single flux rope, and a subsequent oscillatory phase with some ongoing small-scale reconnection. The total kinetic energy of the system initially increases (Figure \ref{fig:energy_evolution}), as the flux ropes accelerate towards each other due to the unbalanced Lorentz forces. A peak is reached around $t = 12$, where a current sheet has formed (see 2\textsuperscript{nd} row of Figure \ref{fig:reconnection_snapshots}), associated with the peak reconnection rate. A peak in the fraction of the domain with current above the critical current and a peak in the the maximum current can also be observed at this time. After this, the kinetic energy, reconnection rate and fraction of the domain with a current above the critical current undergo oscillatory behaviour. This is initially associated with the "bouncing back" of the flux ropes as they partially fail to reconnect \citep{Stanier2013}. The kinetic energy decreases until around  $t = 20$, after which a double peak is observed, and then oscillates with peaks of slowly decaying amplitude through the oscillatory-reconnection phase (shown in Figure \ref{fig:oscillation_snapshots}). Oscillations in the smoothed maximum current of the system are observed to dampen over time, indicating that strength of the current sheets observed in Figures \ref{fig:reconnection_snapshots} and \ref{fig:oscillation_snapshots} also decay over time.\\

The oscillatory nature of the reconnection rate seen in the right-panel of Figure \ref{fig:energy_evolution} is broadly in accordance with \citet{Stanier2013}, but differs in some respects due to the different choices of parameters, particularly our resistivity model. In our case, the reconnection rate actually reverses (e.g. around $t=32$) which will be discussed further below. In the later phases, the reconnection rate more or less saturates at a value of around 0.0012, giving flat tops to the oscillation peaks/troughs. These are associated with the anomalous  resistivity model, as the current (in the absence of any mechanism in this phase to create current sheets), settles around the critical current value.\\

We now consider the evolution of the system in more detail, focusing for now on the case when the background axial magnetic field is $B_0 =2$.

\subsection{Initial Merger Phase}

\begin{figure*}
\makebox[\textwidth][c]{\includegraphics[width=1.2\textwidth]{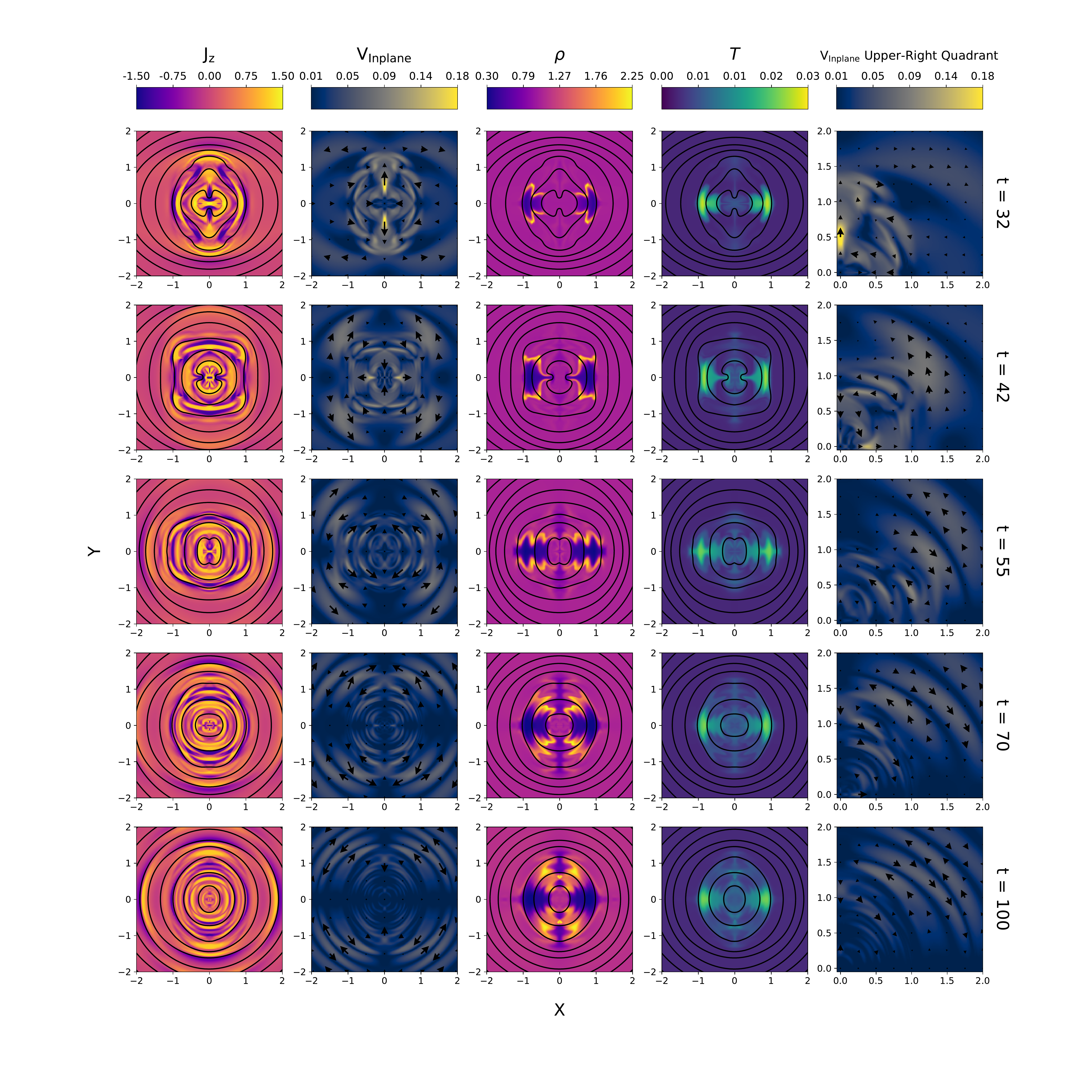}}%
    \caption{Same as in Figure \ref{fig:reconnection_snapshots} but the snapshots are taken at $t = 32, 42, 55, 70, 100$.}
    \label{fig:oscillation_snapshots}
\end{figure*}

The overall evolution of the magnetic field, up to $t = 25$ is shown in Figure \ref{fig:reconnection_snapshots}. To further visualise this, a video of the evolution of the magnetic field between $t = 0 - 100$ (as well as the other plasma parameters discussed in Figures \ref{fig:reconnection_snapshots} and \ref{fig:oscillation_snapshots}) can be found as supplementary material. The initial attractive motion of the two current-carrying magnetic flux ropes can be observed in the snapshots $t = 0$ and $t = 12$. Around  $t = 12$, the two flux ropes meet. At this time, an out-of-plane current sheet ($j_z$) forms near the origin, and the reconnection rate peaks (see Figure \ref{fig:energy_evolution}). The in-plane velocity shows a classic reconnection pattern with  inflows into the current sheet and strong  outflow jets along the x-axis. Subsequently ($t = 16$ and $t = 20$), the flux ropes  continue to undergo strong magnetic reconnection, but with some oscillation of the reconnection rate. The powerful reconnection outflow jets persist, but the development of large-scale vortices is also observed as the jets terminate as they meet the surrounding azimuthal field and azimuthal return flows appear. Finally around $t = 25$, the two flux ropes have completely  merged into a single flux rope, although at this point it is still far from equilibrium. Ongoing reconnection activity after the merging can be observed: a current sheet remains, though it is broader and weaker; characteristic plasma inflows and outflows can still be observed and there is a non-zero reconnection rate at the time (Figure \ref{fig:energy_evolution}).\\

In the early stages of the simulation, the density of the plasma  remains almost constant. Around $t = 16$, the density along the x-axis starts to develop a small depletion, as the plasma flows away from the X-point. At $t = 20$, an increase in the density in the region surrounding the tips of the jets flowing along the x-axis can be observed. This is due to a buildup of plasma as the outflow jets are blocked by the ambient azimuthal magnetic field. \\

Our focus here is not on the energetics; however, the final column of Figure \ref{fig:reconnection_snapshots} shows that the temperature is observed to increase in the region in which the current sheet forms. Thus, as expected, reconnection is associated with plasma heating. Beyond $t = 16$, a cooler region surrounding the tips of the jets flowing along the x-axis can be observed. This reduced temperature region coincides with the higher density region at the jet tips.

\subsection{Oscillatory Reconnection Phase} \label{ref:OscillatoryReconnectionPhase}

By around $t = 25$, the two initial magnetic flux ropes have merged into a single flux rope (see Figure \ref{fig:reconnection_snapshots}). The magnetic field of this flux rope, at this point, is stretched along the x-axis. This orientation will be referred to as a horizontal alignment. Figure \ref{fig:oscillation_snapshots} illustrates how the system continues to evolve, where the newly formed flux rope begins to oscillate. At $t = 32$, the magnetic field has changed from being stretched along the x-axis, to being stretched along the y-axis. This will be referred to as a vertical alignment. The flux rope then proceeds to sweep back to a horizontal alignment between $t = 32$ and $t = 55$. By $t = 70$, the system cycles back to a vertical alignment with the magnetic field lines having a smaller eccentricity than those previously observed. By $t = 100$, the magnetic field is approaching to a circular configuration, though oscillations still persist.\\

\begin{figure*}
\makebox[\textwidth][c]{\includegraphics[width=1.2\textwidth]{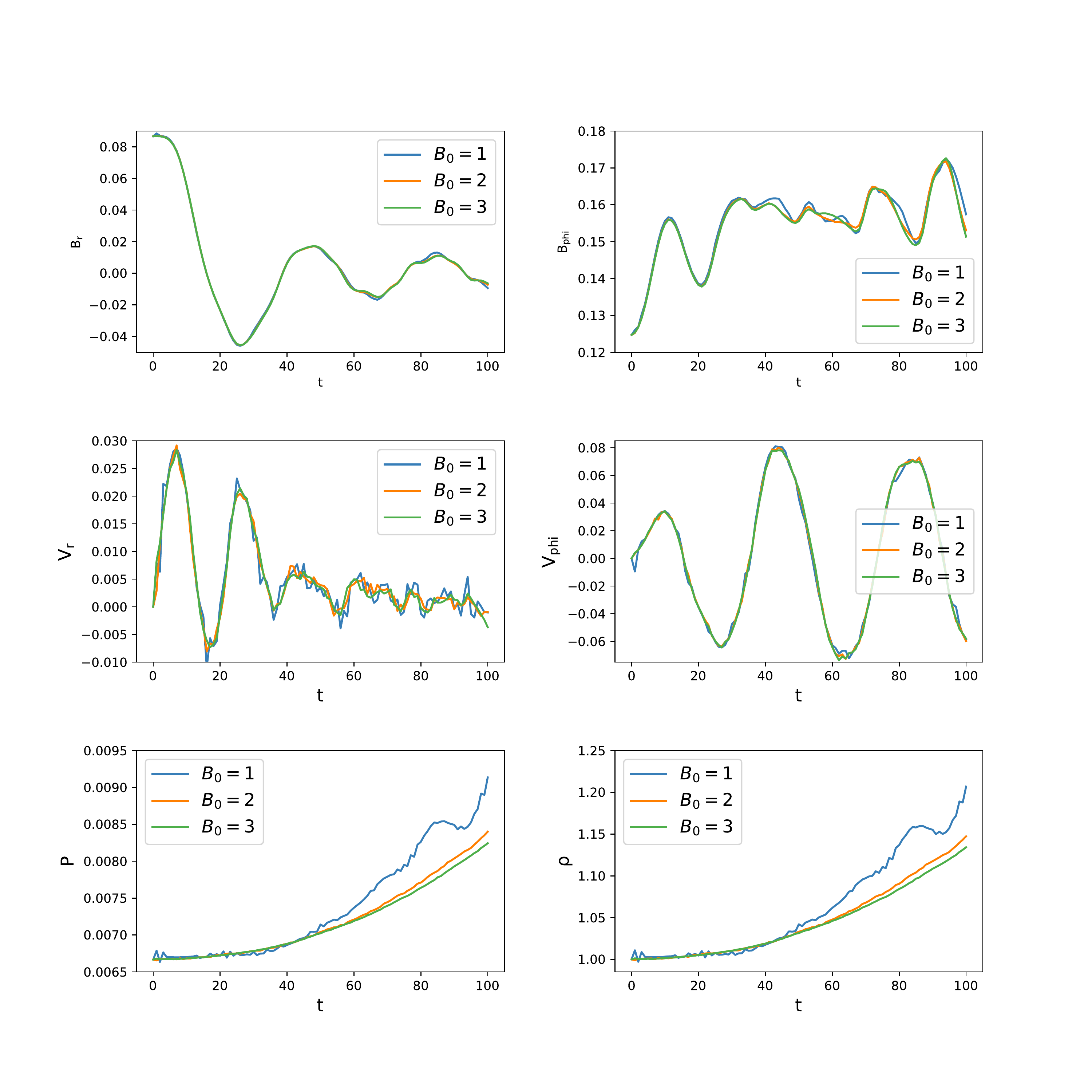}}%
    \caption{The variation of in-plane magnetic field $B_r$, $B_{\phi}$, in-plane velocity $V_r$, $V_{\phi}$, pressure and density with $t$ at $r = 1.8$ on the $y = x$ axis is shown for $B_0$ = 1, 2, 3.}
    \label{fig:temporal_fluctuations}
\end{figure*}

It can also be observed that at $t = 32$, plasma now flows away from the reconnection site along the vertical axis, while flowing inwards along the x-axis. This is associated with a short vertical current sheet between what are likely to be two small magnetic islands to the left and right of the origin. Note from Figure \ref{fig:temporal_fluctuations} that the reconnection rate has reversed at this time, consistent with the changed sign of $j_z$ at the origin. Outside of the reconnection site, the presence of plasma flowing in the azimuthal direction can be also be observed. By $t = 42$, the plasma flow along the vertical axis has switched direction, now flowing inwards with outflows along the horizontal axis as in the earlier phases, but now weaker.  Plasma has now also begun flowing along the azimuthal direction away from the x-axis towards the y-axis creating two large-scale vortices. By $t = 70$, the system returns to having an outflow along the vertical axis away from the origin, however the magnitude of the velocity of this outflow is greatly reduced when compared to previously observed outflows. Plasma closer to the origin can be observed flowing along the azimuthal direction away from the y-axis towards the x-axis, the opposite direction to that observed at $t = 42$. The azimuthal flow at $t = 42$ can still be observed at $t = 70$ but shifted radially outwards away from the reconnection site. By $t = 100$, reconnection outflows/inflows can no longer be observed but flows in the azimuthal direction have continued to develop between $t = 70$ and $t = 100$. At $t = 100$, successive wavefronts of azimuthal flow are observed propagating radially outwards away from the reconnection, concentrated  along each diagonal axis. These propagating wavefronts are perhaps indicative of MHD waves that can be associated with QPPs. We discuss the nature of these waves further in Section \ref{PhysicalPropertiesofWaves} below, noting that it is not expected that they should correspond to any single MHD wave mode. This is because the background field is highly inhomogeneous (with strong variations in wave speed within one wavelength, as well as the equilibrium field lines having significant curvature), and the disturbances have a high amplitude and are therefore nonlinear.\\

Finally, referring back to Figure \ref{fig:oscillation_snapshots}, the density and temperature of the plasma can be observed between $t = 32$ and $t = 100$. The low density, high temperature region along the x-axis with a higher density, lower temperature envelope, first observed to form in Figure \ref{fig:reconnection_snapshots} continues to develop as the system evolves. Though density and temperature also varies along the vertical axis during this time, their contributions are less than those made during the initial reconnection phase. The fluctuations in density and temperature are not observed in the region in which the emitted wavefronts propagate which is an important property for understanding what physically drives these waves.

\begin{figure*}
	\makebox[\textwidth][c]{\includegraphics[width=1.2\textwidth]{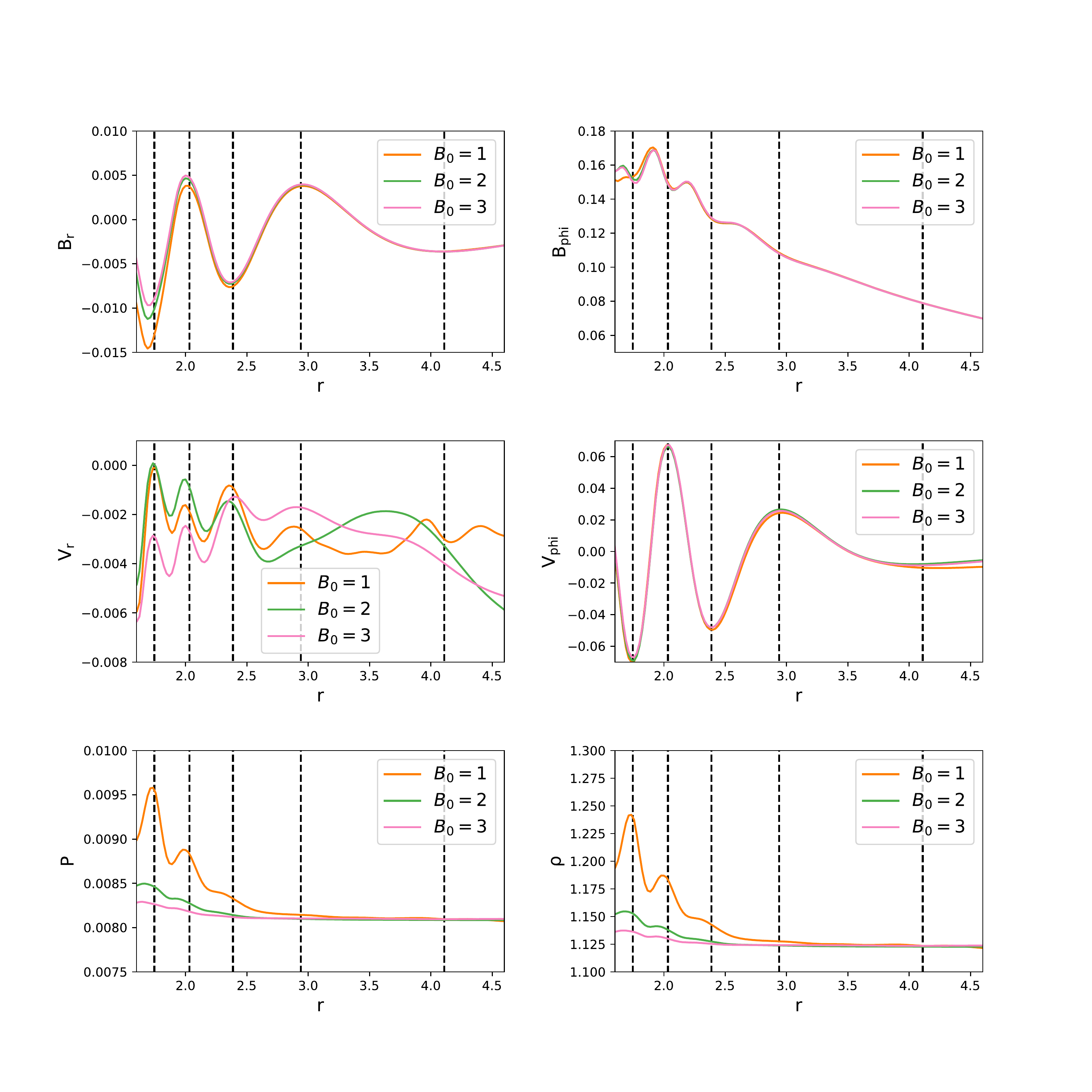}}%
    \caption{The variation of $B_r$, $B_{\phi}$, $V_r$, $V_{\phi}$, $P$ and $\rho$ with $r$ at $t = 100$ is shown for $B_0 = 1, 2, 3$. Data has been taken along the $y = x$ axis, where $r > 1.6$, in the direction of one of the emitted wavefronts. The positions of the wavefronts associated with azimuthal velocity $V_{\phi}$ are marked with dashed black lines.}
    \label{fig:diagonal_fluctuations}
\end{figure*}

\begin{figure*}
\makebox[\textwidth][c]{\includegraphics[width=1.2\textwidth]{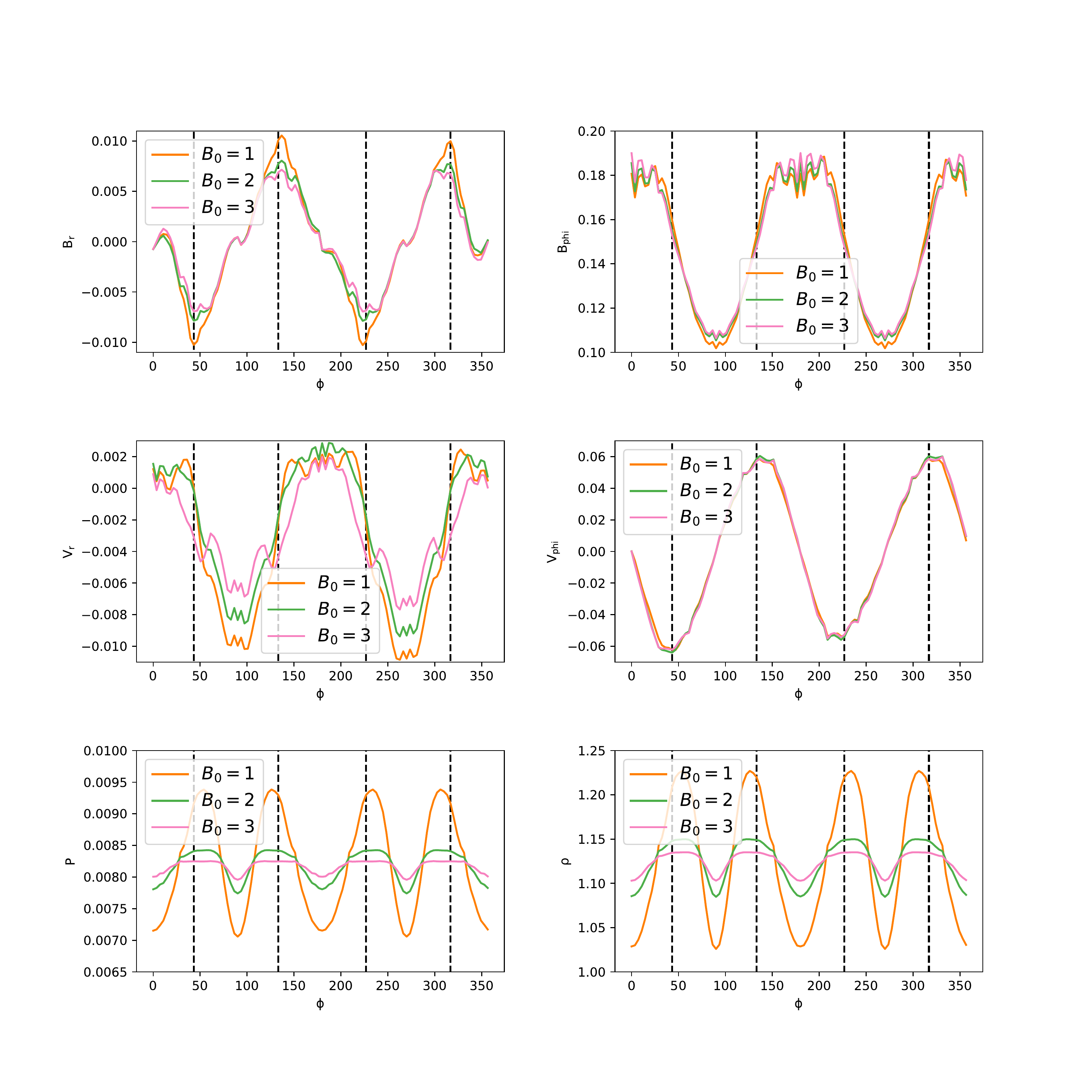}}%
    \caption{The variation of $B_r$, $B_{\phi}$, $V_r$, $V_{\phi}$, $P$ and $\rho$ with $\phi$ at $t = 100$ is shown for $B_0 = 1, 2, 3$. Here, $B$ and $V$ represent the in-plane magnetic field strength and the in-plane velocity respectively. Data is taken along a constant radius of $r = 1.8$, the position of an emitted wavefront along the $y = x$ axis. The positions of each emitted wavefront around the reconnection site at $t = 100$ are marked with dashed black lines.}
    \label{fig:circular_fluctuations}
\end{figure*}

\subsection{Properties of the Emitted Waves}\label{PhysicalPropertiesofWaves}

We have seen that there are distinctive  outward propagating ``ripples" in the in-plane velocity with quadrupolar structure and with maximum amplitude along the diagonals $y = \pm x$ and with flow predominantly in the azimuthal direction (see second column of Figure \ref{fig:oscillation_snapshots}, particularly from about t = 55 onwards). We now investigate these in more detail. These ripples propagate predominantly perpendicular to the magnetic field; however the propagation speed is far too low for them to be fast magnetosonic waves. Furthermore, the symmetry of our 2D model prohibits any waves from propagating along the dominant axial field component ($B_z$) and this field component can play no role in any waves (since $\mathbf{k} \cdot B_z\mathbf{\hat{z}}=0$). We thus hypothesised that these ripples are governed only by the in-plane magnetic field. This was tested by running the simulations for both increased and reduced values of $B_z$. Therefore all results in Figure \ref{fig:temporal_fluctuations}, \ref{fig:diagonal_fluctuations} and \ref{fig:circular_fluctuations} are shown for  $B_0 = 1,2,3$.\\

To understand the ``ripple" structures better, we first  plot the variation of $B_r$, $B_{\phi}$, $V_r$, $V_{\phi}$, $P$ and $\rho$ with $t$, at a fixed point $r = 1.8$ along the $y = x$ line (Figure \ref{fig:temporal_fluctuations}). We select this location because it is evident from Figure \ref{fig:oscillation_snapshots} that the perturbations are largest  along the diagonals. Oscillations are observed in both magnetic field and velocity, with $V_{\phi}$ dominating $V_r$ contributions. The oscillations in radial velocity are  lower in amplitude (by a factor of about 4.5), are noticeably noisier and are  dominated by a component with twice the frequency of the main $V_{\phi}$ oscillation. A similar relationship is observed between the dominant $B_r$ oscillation and the noisier $B_{\phi}$ oscillation. Note that the frequency of the main mode (in $V_{\phi}$ and $B_r$) corresponds to the oscillating reconnection patterns seen in Figure \ref{fig:oscillation_snapshots} which are discussed above. Pressure and density increase over time but oscillations in these variables are not discernible. The time variations of magnetic field and velocity  are almost identical for the different  values of $B_0$, particularly for the main oscillations seen in radial magnetic field and azimuthal velocity. However, the magnitude of the guide field ($B_0$) does have some effect on the pressure and density variations. This is because the stronger axial fields constrain the in-plane plasma motions more, leading to weaker variations in density and pressure. \\

The properties of the emitted waves are further explored in Figure \ref{fig:diagonal_fluctuations} and Figure \ref{fig:circular_fluctuations} where we show the spatial variations at a fixed time ($t = 100$). Figure \ref{fig:diagonal_fluctuations} illustrates how $B_r$, $B_{\phi}$, $V_r$, $V_{\phi}$, $P$ and $\rho$ varies with $r$ along the $y = x$ line for $r > 1.6$. focusing on behaviour outside of the reconnection site. Figure \ref{fig:circular_fluctuations} illustrates how these same variables vary with $\phi$ around the circle of  radius $r = 1.8$; at $t = 100$, this radius  is at the peak of one of the radially-propagating wavefronts.\\

Considering first radial variations (Figure \ref{fig:diagonal_fluctuations}), the maxima and minima of the $B_r$ and $V_{\phi}$ oscillations are observed to be in-phase. Similar behaviour is observed in the variation of these variables with respect to azimuthal angle  $\phi$ (Figure \ref{fig:circular_fluctuations}), with the  maxima and minima of the oscillations in $B_r$ and $V_{\phi}$ being  aligned.  This shows the dominant mode has a  $m=2$ structure (where $m$ is the azimuthal mode number), which is also evident in Figure \ref{fig:oscillation_snapshots}. The other field and velocity components also have clear $m=2$ structure, although there are significant higher mode-number components in the $V_r$ signal in particular. Whilst the spatial  waveforms  for $B_{phi}$, $B_r$ and $V_{phi}$ are all almost independent of the axial field $B_0$, the radial velocity becomes generally weaker as $B_0$ increases: this may be due to the guide field suppressing reconnection outflows. \\

The relationship between $B_{\phi}$ and $V_{r}$ oscillations is a little more complex, but at $t = 100$ they are clearly correlated up to about $r = 3.0$ (Figure \ref{fig:diagonal_fluctuations}). The minima of $B_{\phi}$ and the maxima of $V_{r}$ with respect to $r$ are observed to occur at each wavefront between $r = 1.6$ and $r = 3.0$. These variables are observed to share a wavelength that is half of that of $B_r$ and $V_{\phi}$. At larger radii, the oscillatory behaviour is less clear in both these quantities, and their is less correlation between them. This region is far from the reconnection site and the ``ripple" structure is much weaker.

In Figure \ref{fig:diagonal_fluctuations}, the pressure and density of the plasma is observed to decay as the radius away from the origin increases. Both variables exhibit oscillations which are much lower in amplitude than those in $\bf{B}$ and $\bf{v}$ and have maxima that are in-line with the location of the emitted wavefronts. However, the radial  wavelength of these variations is half that of the magnetic field and  velocity oscillations. Similarly, in Figure \ref{fig:circular_fluctuations}, the pressure and density oscillations with respect to $\phi$ are also observed to have maxima in-line with the velocity wavefronts but with half the wavelength. The minima of the density and pressure oscillations are located along the x and y-axis. This is due to the azimuthal velocity moving plasma away from the x and y-axis, into the wavefronts along the diagonal axis. Note also that the spatial structure of the pressure and density fluctuations is significantly affected by the guide field.  \\

\begin{figure*}
\makebox[\textwidth][c]{\includegraphics[width=1.0\textwidth]{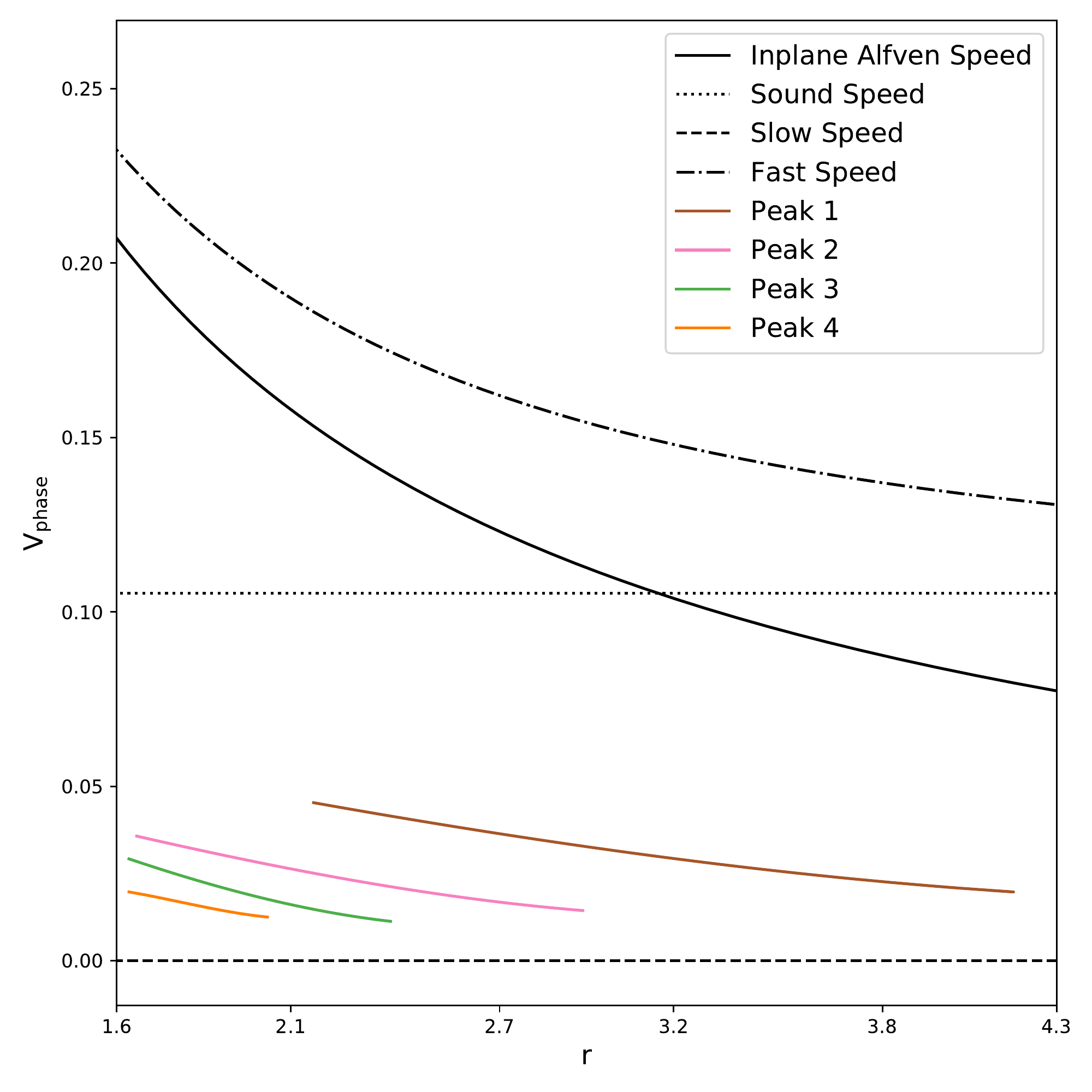}}%
    \caption{The variation of theoretical values for linear MHD wave modes, including in-plane Alfv\'en speed, sound speed, fast and slow magnetoacoustic wave speed with $r$ are calculated at $t = 100$ and plotted for $r > 1.6$. The slow and fast magnetoacoustic waves are assumed to be propagating perpendicular to the magnetic field. The observed phase speed of each emitted wavefront in the simulation and their variation with $r$ for $r > 1.6$ is also plotted.
}
    \label{fig:vphase_analysis}
\end{figure*}



\subsection{Phase Speed Analysis of the Emitted Wavefronts}

It is possible to determine more about the emitted waves by analysing their phase speed and comparing them with known linear MHD modes (see Figure \ref{fig:vphase_analysis}). In this figure, the phase speed calculated for  each peak  of the propagating waves ($r_{peak}$) along the $y = x$ line, for $x > 0$ is plotted as a function of $r$ for the region  $r > 1.6$ where the outward-propagating ripples are evident. To calculate the phase speed, the position of the peak amplitude $r_{peak}$ was first tracked for each wavefront  between $t = 32$ and $t = 100$. With this data, the phase speed of each maxima ($v_{phase}$) for each time-step was determined by calculating $v_{phase} = \frac{\Delta r_{peak}}{\Delta t}$. By knowing the position of the maxima as a function of time, it was then possible to determine the variation of $v_{phase}$ for each maxima as a function of $r$. Peak 1 in Figure \ref{fig:vphase_analysis} is the earliest wavefront to be emitted, and is thus also the outer-most travelling wavefront. Each successive numbered peak corresponds to each successively emitted wavefront. It is observed that the phase speed of each wavefront decays as it propagates radially outwards. Each successive emitted wavefront is also observed to have with a lower speed than the previous wavefront. As the amplitude of the oscillatory reconnection decays over time, and the reconnection site has less affect on plasma far away from it, these observations suggest the the phase speed of the emitted waves is linked to the reconnection process.\\

Figure \ref{fig:vphase_analysis} also shows for reference the values for the local phase speeds of linear MHD wave  modes in a straight uniform field: the in-plane Alf\'en speed, sound speed, and fast and slow magnetoacoustic wave speed.  The in-plane Alfv\'en speed is defined as the Alfv\'en speed based only on the local value of the equilibrium field $B_{\phi}$ and ignoring $B_z$, on the basis that, as seen above, the  observed waves are almost independent  of $B_0$. Since the observed waves propagate almost exactly perpendicular to the local mean in-plane field, the fast and slow speeds are plotted for perpendicular propagation. Our system is varying in both space and time, with both long-term/large-scale trends  and  rapid/small-scale fluctuations, which cannot easily be separated. Here, we identify the background equilibrium state as being a single circular flux rope with constant density, and calculate the wave speeds based on this. Of course, it must be borne in mind these values are only applicable for straight uniform fields, but the local values are still indicative.  It can be seen that the Alfv\'en speed varies significantly  across the region of interest, and there is a transition from a magnetically-dominated region ($\beta_i < 1$) closer to the reconnection site  to a pressure-dominated region ($\beta_i > 1$) at larger radii; noting that the in-plane $\beta$ value $\beta_i$ is much lower than the total $\beta$ (including the axial field $B_0$) which is everywhere small. The phase speeds decline for successive wave peaks, suggesting each wavefront is emitted with somewhat lower phase speed than its predecessors, and the phase speed for each wavefront  also falls with increasing radius, following qualitatively the profiles of the Alfv\'en and fast speed, but being significant lower than either.

\section{Discussion}\label{Discussion}

The two initial flux ropes attract each other due to the Lorentz force and collide, beginning the process of oscillatory reconnection. As the flux ropes approach each other, a current sheet forms between the two flux ropes, along the x-axis, allowing reconnection to begin. Characteristic reconnection outflow jets along the x-axis are initially seen. This creates a heated, less dense environment in the vicinity of the jets, with cooler, denser plasma envelopes surrounding the region. Around $t = 12$, the reconnection rate begins to decrease towards a local minimum  characteristic of the system entering a ``sloshing regime", as described by \citet{Knoll2006}: when two magnetic islands collide in an  environment with a high Lundquist number, there is a reduction in the reconnection rate due to a significant buildup of magnetic pressure near  current sheet, opposing the coalescence process. This phenomenon is also observed in \citet{Stanier2013}'s simulations. After around 25 Alfv\'en times, the two flux ropes have merged into a single flux rope - though with considerable sub-structure, and still far from equilibrium.\\

The magnetic field at this time is in a ``horizontally-aligned" configuration, in which the in-plane magnetic island is extended along the $x$-axis. In this alignment, the central current sheet is also oriented horizontally, and some reconnection continues,  creating a pair of small islands either side of the X-point along the $x$-axis. This may be seen as the reconnection "over-shooting", as the field has reconnected more than is needed to form a single flux rope. The field then continues to evolve - in the first cycle of vertical alignment ($t=32$),   the central current sheet has reversed (compared with earlier times), and the two small central magnetic islands are now merging; the reversed sign of $j_z$ at the origin, and hence the reconnection rate, corresponds to the fact the the current is related to the oppositely-directed $B_y$ field components across the origin.    Subsequently, the system enters an oscillatory relaxation phase, where the magnetic field structure  cycles between horizontal and vertical alignments, with initially  strong  - but decaying - distortions from the equilibrium circular field lines. The  orientation of the central current sheet, and the direction of the axial current, as well as the associated inflow/outflow structure, similarly oscillates.\\

The cycle repeats, but decreasing in amplitude, until the end of our simulation ($t = 100$), when  the magnetic field is close to a circular, equilibrium state, although some  flows persist, and it would take more time for all activity to be damped away. During this oscillatory  relaxation phase, reconnection outflow jets form parallel to the alignment of the magnetic island, and so the orientation of the inflow/outflow flow system also oscillates between horizontal (as in the initial merging phase) and vertical alignments, while the  speed of these jets reduces with each cycle. The outflow jets are largely blocked by the ambient azimuthal magnetic field, and thus  plasma from the outflows moves in the azimuthal direction around the reconnection site, creating a large-scale vorticial flow system with strong azimuthal flows connecting the instantaneous outflows and inflows. As the magnetic island orientation cycles, this flow system reverses.  The region in which the jets form contain hotter, less dense plasma than its environment, leading to a pattern of density depletions mainly along the $x$-axis (mainly) and also more weakly along the $y$-axis. \\ 

During the relaxation phase, the reconnection rate at the origin oscillates around zero, as discussed above. The dominant contribution to the reconnection rate is $-j_z$ suggesting that these oscillations are due to the creation and dissipation of current sheets at the origin. The  reconnection rate damps over time due to the reduced magnitude of the current sheets as the system continued to relax. The maxima/minima of the reconnection rate oscillations saturate at around 0.0012. This value is consistent with $\eta = 10^{-3}$ multiplied by the critical current $j = 1.2$, which is the dominant contribution to the reconnection rate calculation. Whilst in the early phase of driven reconnection, the current can - and does - rise above the critical value (and indeed, in common with other reconnection simulations, this allows reconnection to occur), in the later phase we have  more distributed currents with no forcing of  current sheet formation. Thus, as the current reaches its critical value, anomalous  resistivity sets in and the current does not grow further. \\

The sweeping of the magnetic field between horizontal and vertical alignments, and the generation of associated azimuthal flows described above, leads to  ``ripples" with successively reversing peaks of  azimuthal velocity  forming  and propagating  away from the reconnection region. These may be considered as waves on a background structure of circular magnetic field lines.
Oscillations in $B_r$, $B_{\phi}$, $V_r$ and $V_{\phi}$ are found within the region in which the ripples  propagate.\\

The dominant mode of oscillation is exhibited by $B_r$ and $V_{\phi}$. $B_{\phi}$ and $V_{r}$ are also found to share a frequency and a wavelength in the radial direction at $t = 100$ that is half of the wavelength of the main oscillation.  The oscillations in $B_{\phi}$ and $V_r$ are observed to be aligned with the wavefronts for $1.6 < r \lessapprox 3.0$ at $t = 100$. However for $r \gtrapprox  3.0$ this property ceases to be true. This implies that the oscillations are connected to a process close to the reconnection site. Furthermore, moving in the azimuthal direction at $r = 1.8$ at $t = 100$, the maxima of the oscillations of these variables are found to align with 0\textdegree, 90\textdegree, 180\textdegree and 270\textdegree, the region in which jets are emitted along the x and y-axis. Finally, $V_r$ and $B_{\phi}$ oscillations are found to dampen with respect to $r$ over time, as are the magnitude of the velocity of the emitted jets. Thus it is believed that the $V_r$ and $B_{\phi}$ oscillations originate from the jet outflows and dampen over time as the reconnection site continue to relax.\\

Overall, the "ripples" appear to consist of a standing wave along the circular field lines, involving parallel flows and oscillation in the radial magnetic field associated with the distortions of the field lines, with also slow outward propagation. The phase speed of the outward propagating disturbances is found to be small compared with the local values of the sound and Alfv\'en speeds, although the declining speed with radial distance does reflect the Alfv\'en speed variation.  The lack of significant pressure perturbations suggests that these are not sound or slow  magnetoacoustic waves, and it appears that the pressure/density variations with  doubled frequency  arise as a weaker secondary effect through nonlinear wave coupling of the main, predominantly magnetic, mode.  However, while propagation perpendicular to the equilibrium field suggests a fast wave, the very much slower phase speed seems to rule this out.
 The  amplitude of the waves is large (velocities up to 8\% of the  Alfv\'en speed, magnetic field fluctuations similarly around 8\% of the background in-plane field) so they are highly nonlinear. Furthermore, the waves are propagating on a background which is highly non-uniform, and cannot be considered using standard approximations of a slowly-varying medium (e.g WKB): firstly, the equilibrium field is circular, with radius of curvature comparable with the wavelengths (both azimuthal and radial) of the waves, and secondly,  the equilibrium field magnitude also depends strongly on radius (with  $1/r$ dependence outside the current-carrying flux rope). In such a context, it is not  meaningful to identify waves as corresponding to any single standard MHD wave mode.  In some respects, the properties resembles the "superslow" apparent propagation of wavefronts associated with phase-mixing, identified by \citet{Kaneko15}, which is found  not be a true wave propagation. However, this phenomenon arises in a system with shear  Alfv\'en waves on adjacent field lines which phase-mix due to the variations in Alfv\'en speed, and the disturbances are in the out-of-plane direction. By contrast in our case, the oscillations are in-plane, and the frequency is determined by the oscillatory reconnection rather than corresponding to the Alfv\'en frequency of each field line. \\

As stated earlier, two examples of flux ropes merging in flares are merging coronal loops and merging plasmoids in fragmented current sheets. Using the observed values of the emitted waves presented in this paper, it is possible to estimate physical parameters of the MHD waves emitted in these scenarios. In dimensionless units, the observed wavelength and frequency of the radially propagating wavefronts are found to be 1 and 0.025 respectively. Considering a coronal loop to be of radius $1000$ km, with a background magnetic field on the order of $5*10^{-3}$ T, in a plasma of density $10^{9}$ cm$^{-3}$, it can be calculated that the emitted waves would have a wavelength of order $1000$ km, frequency $0.9$ Hz and an approximate phase speed of 90 km/s. Similarly, considering  a plasmoid in a fragmented current sheet  of length scale $10$ km, with a background magnetic field on the order of $5*10^{-3}$ T, in a plasma of density $10^{9}$ cm$^{-3}$, it can be calculated that the emitted waves would have a wavelength of order $10$ km, frequency $90$ Hz and an approximate phase speed of 900 km/s.\\

Conversely, observing the frequency and wavelength of emitted MHD waves formed during flux rope coalescence would allow  estimation of the background magnetic field of the system and the site of the structures involved. This, along with the strong link between the frequency of the emitted MHD waves and the frequency of the oscillatory reconnection suggests that studying properties of the emitted MHD waves formed during flux rope coalescence could provide insight and be used as a diagnostic tool to study a coalescing region.

\section{Conclusions}

In this paper, we have presented 2D MHD simulations of two current-carrying cylindrical flux ropes merging in a coronal plasma using the LARE2D code. The flux ropes mutually attract and merge into a single flux rope through a process of oscillatory reconnection. After the single flux rope is formed, it continues to oscillate strongly for some tens of Alfv\'en times, cycling between elliptically-distorted configurations. Strong perturbations in magnetic field and velocity are evident, with quadrupolar structure, while much weaker density and pressure perturbations with shorter wavelengths and higher frequencies are likely associated with mode conversion. During the later stages of oscillatory reconnection, large-amplitude waves with phase speed unexpectedly low compared to linear MHD wave modes in a uniform field  are emitted radially with peak amplitude focused along each diagonal.\\

The emitted waves contain strong azimuthal velocities that alternate direction between each successive wavefront, associated with in-phase variations in the radial magnetic field; there are also outward propagating disturbances in the radial velocity and azimuthal field, with lower amplitude. The general behaviour of the emitted waves was found to be almost independent of the background out-of-plane magnetic field, which is expected since within our 2D model (representing long flux ropes), wave propagation along the flux rope is not possible. The waves are both strongly non-linear and propagate in a highly non-uniform background field, and so cannot be identified simply with any single standard MHD mode. \\

From the results presented in this paper, it can be concluded that flux rope coalescence in the solar corona intrinsically leads to oscillatory reconnection. Furthermore the process leads to the production of wave-like disturbances propagating radially outwards away the flux rope, which are highly non-linear and are strongly affected by the non-uniform background field, and thus do not correspond to any single MHD waves mode.  The frequency of the oscillations, the wavelength and the phase speed  of the propagating disturbances could be used as diagnostic tool to determine the magnetic field in a region undergoing flux rope coalescence. Furthermore, the outward propagating waves  may also play a significant role in energy transport in flaring plasmas.\\ 

Our model is quite generic, and the key finding is to   demonstrate conclusively the principle that oscillations and waves can arise naturally in a reconnecting magnetic field, without any external oscillatory driver, as well as  to reveal in some detail  the mechanisms by which such oscillatory behaviour arises. A natural future development would be to develop more realistic models of reconnecting coronal fields, using 3D MHD simulations, and to forward-model the emission. Indeed, \citet{Smith2022} have recently shown that oscillations arise in a flaring twisted loop, and that  microwave oscillations are produced which are comparable with QPPs. However, our simpler model can be considered to represent two scenarios relevant to flares - large-scale loop interactions and plasmoid coalescence within looptop current sheets -  and as it is a pure MHD model, the predicted frequencies etc. can be re-scaled to match different field strengths and length scales (our predicted frequencies scale with the Alfv\'en wave travel time).\\

It is estimated that MHD waves emitted through merging coronal loops would have a wavelength of order $1000$ km, frequency $0.9$ Hz and an approximate phase speed of $90$ km/s, whereas  MHD waves emitted through merging plasmoids in a fragmented current sheet would have shorter wavelengths and higher  frequencies, commensurate with the plasmoid size. Thus, the oscillations could be used to provide diagnostic information about the reconnection process and the sizes and magnetic field strengths of merging flux ropes.

\section*{Acknowledgements}

We wish to thank the Science and Technology Facilities Council (STFC) for providing studentship support for JS. PKB and MG are funded by the STFC grant ST/T00035X/1. We thank the Distributed Research utilising Advanced Computing (DiRAC) group for providing the computational facilities used to run the simulations in this paper. We would like to express our gratitude to Ineke De Moortel for insightful discussions on possible interpretations of the results presented in this paper. We are also grateful to Tom Van Doorsselaere for drawing our attention to superslow wave propagation. 

\section*{Data Availability}

The data underlying this article will be shared on reasonable request to the corresponding author (JS).



\bibliographystyle{mnras}
\bibliography{bibliography} 








\bsp	
\label{lastpage}
\end{document}